\documentclass{article}
\usepackage[T1]{fontenc}
\usepackage[utf8]{inputenc}
\usepackage{ismir}
\usepackage{amsmath,cite,url}
\usepackage{graphicx}
\usepackage{color}
\usepackage{multirow}
\usepackage{booktabs}
\usepackage{caption}
\usepackage{subcaption}

\title{Spacing Out: On the Reliability of Binaural\\Music Source Separation Metrics}

\multauthor
  {Richa Namballa$^1$ \hspace{1cm} Magdalena Fuentes$^{1, 2}$}
  {
  $^1$ Music and Audio Research Laboratory, New York University, United States\\
  $^2$ Integrated Design and Media, New York University, United States\\
  {\tt \small richa.namballa@nyu.edu}
  }

\def\authorname{R. Namballa and M. Fuentes}

\begin{document}

\maketitle

\begin{abstract}
\vspace{-6pt}
Despite the rising popularity of immersive audio, binaural music remains underexplored in music information retrieval (MIR), particularly regarding the task of music source separation (MSS).
While existing stereo MSS models can process binaural audio, they often degrade the spatial quality of the separated stems and undermine listener immersion.
Through a perceptual study comparing binaural and stereo MSS outputs, we evaluate how well objective spatial distortion metrics correlate with human perception.
Our findings reveal varied agreement between these metrics and human judgment, highlighting a lack of reliability when used to evaluate binaural music tasks.
Specifically, we find that Interaural Time Difference (ITD) estimation is highly sensitive to noise and separation artifacts.
In evaluating two alternative ITD estimation methods, we uncover a critical trade-off between robustness and accuracy, particularly for narrow-band instruments like bass.
These results underscore the need for accurate, interpretable spatial metrics designed for binaural music to  develop models that preserve source localization and listener immersion.
\end{abstract}

\vspace{-8pt}
\section{Introduction}\label{sec:introduction}

Music source separation (MSS) is the task of separating a musical mixture into its corresponding stems (sources).
Recent advancements have resulted in impressive separation quality for stereo recordings~\cite{Araki2025}.
However, little work has been done in binaural MSS, despite the growing interest in immersive sound in the audio community~\cite{Namballa2025, Kasak2023}.

Binaural audio is a popular choice for rendering spatial audio in virtual and augmented reality (VR and AR) systems.
Its efficient processing combined with its seamless headphone compatibility make it an ideal type of immersive audio to investigate further in a musical context.
Events such as concerts could benefit from binaural capture and MSS to deliver well-mixed performances to virtual attendees.
Prior research has also shown the effectiveness of binaural source separation on environmental sound~\cite{Veluri2023} with important applications in accessibility.
Expanding this task to music allows listeners with different hearing abilities or neurodivergence to enjoy musical performances without sacrificing the immersive experiences which make these live events special.

A recent study found that existing stereo music source separation models often fail to preserve spatial cues in binaural mixtures~\cite{Namballa2025}.
However, this conclusion relies on spatial metrics whose perceptual relevance and robustness have not been systematically validated for binaural music source separation. In this work, we investigate how well spatial metrics align with human judgment and how robust they are to separation artifacts.

To facilitate our study, we retrain an existing MSS architecture on binaural music data and compare it against its stereo-trained counterpart.
We then conduct a perceptual study in which listeners evaluate separated stems from a binaural mixture in terms of overall quality and localization ability.
Our results show that the Signal to Residual Distortion Ratio (SRR) and change in interaural level difference ($\Delta$ILD) metrics agree most consistently with listener preference, while the change in interaural time difference ($\Delta$ITD) shows weaker agreement.
This discrepancy depends on both the instrument's position of placement and type.
We find that agreement is weaker for centered sources, where listeners also struggle to distinguish stereo- and binaural-separated stems, and that $\Delta$ITD is especially unreliable for narrow-band sources, such as bass.

We further analyze $\Delta$ITD and find that its low perceptual agreement is largely due to the fragility of the cross-correlation calculation of the estimated stems' ITD.
Minor separation artifacts and residual interference can cause ITD estimates to collapse.
Despite exploring several mitigation strategies, these artifact-induced failures persist.
Our findings suggest that $\Delta$ILD and SRR are currently more reliable for evaluating binaural MSS, but they primarily reflect loudness balance and residual interference rather than spatial position preservation.
These shortcomings highlight the need for spatial metrics that are both perceptually-grounded and robust to the artifacts introduced by music source separation in both stereo and binaural settings.

\vspace{-10pt}
\section{Related Work}\label{sec:related_work}

The primary difference between binaural and stereo audio is that binaural signals have been filtered through specific time, intensity, and spectral information to mimic human localization cues~\cite{Roginska2017}.
The perception of binaural audio heavily relies on the complex interaction between ITD and ILD cues~\cite{Rayleigh1907} as they allow listeners to identify the direction a sound is originating from.
Even minor discrepancies in these cues could change the perceived location of a sound so it is imperative that these nuances are protected in signal processing systems to maintain a convincing spatial auditory image for the listener.
Therefore, when we evaluate binaural MSS models, we hope that the spatial metrics accurately reflect what is preserved versus what is lost in the separation process both in terms of intensity and timing.

Binaural source separation is well-established in the signal processing research community under various terms, the most  recent being target sound extraction~\cite{Ochiai2020, Delcroix2020, Delcroix2021, Veluri2023TSE}.
The spatial form of this task aims to isolate a specific sound (often speech) from an environment or audio mixture while also preserving the positioning of the source in space; it simultaneously separates and localizes sources.
This line of work has been invaluable in designing deep learning architectures which preserve binaural information in their outputs ~\cite{Tokala2024}.
However, while many implementations work to isolate a general ``music'' class, little work has been done to separate individual instruments from a binaural musical mixture\cite{Veluri2023}, which motivated a recent work on binaural MSS~\cite{Namballa2025}. 

The authors of \cite{Namballa2025} previously released a dataset of binaural music synthesized from the popular MSS dataset MUSDB18-HQ~\cite{Rafii2019} and a database of head-related transfer functions (HRTFs)~\cite{Armstrong2018}; this dataset is known as Binaural-MUSDB.
In their work, they evaluated three well-known MSS models on the test sets of both Binaural-MUSDB and MUSDB18-HQ (Stereo-MUSDB) using metrics based on energy ratios and interaural cues.
In particular, they used the task-agnostic Signal to Spatial Distortion Ratio (SSR)~\cite{Watcharasupat2024}, which quantifies the amount of spatial error introduced to a signal by modeling ITD and ILD changes in a single value, reflective of the duplex theory of localization~\cite{Rayleigh1907}.
This decomposition of error in the estimated signal isolates the spatial error from the non-spatial one, which is reflected in a different energy ratio accounting for residual error, known as SRR.
In parallel, the creators of the Semantic Hearing real-time binaural sound extraction wearable~\cite{Veluri2023} evaluated their model by measuring improvements in the scale-invariant Signal-to-Noise Ratio (SI-SNR)~\cite{LeRoux2019} and the change in the interaural time and level differences ($\Delta$ITD, $\Delta$ILD) between the reference and estimated sources.
These metrics were also included in \cite{Namballa2025}.
As some of these metrics originate from adjacent domains such as speech localization, their utility and perceptual relevance for separated music stems remains unclear.
For example, studies have shown that the accuracy of time difference of arrival (TDOA) estimates, used for ITD calculations, are sensitive to diffuse noise and reverberation~\cite{Lee2020} and even to the bandwidth of a source~\cite{Clifford2013}.
Thus, researchers have explored modifications to the Generalized Cross-Correlation with Phase Transform (GCC-PHAT)~\cite{Knapp1976} method to improve the calculation's robustness, especially in the context of speech~\cite{Donohue2007, Lee2020}.

Lastly, it has not been studied how well the metrics for binaural MSS evaluation correlate with perceptual quality.
Since binaural audio is designed for immersive human listening, it is important to address whether the metrics we use for performance assessment translate into an improved listener experience in the context of music separation.

\vspace{-10pt}
\section{Method}\label{sec:method}

We study how well the spatial metrics for evaluating binaural MSS reflect listener judgments and how robust they are to artifacts in separated music stems.
First, we retrain an existing MSS architecture on binaural music data and compare it with its pretrained stereo counterpart, allowing us to test whether the metrics reflect improvements from in-domain training.
Second, we conduct a perceptual study in which listeners compare the separated outputs in terms of quality and localization ability, enabling us to measure how well each metric aligns with human judgment across source positions and instrument types.
Finally, motivated by the weak perceptual agreement of $\Delta$ITD, we analyze its implementation and stress-test GCC-PHAT-based ITD estimations under varying noise conditions and artifacts.

\vspace{-8pt}
\subsection{Binaural and Stereo Models}\label{subsec:training}

We retrain an MSS model on the training set of Binaural-MUSDB, which follows the original train-test split of Stereo-MUSDB.
Within the training set, we set aside 10 tracks for validation. All tracks are sampled at 44.1 kHz, and the model is trained on 11-second audio segments.

We train our binaural model using the small version of the Sparse Compression Network (SCNet)~\cite{Tong2024} architecture due to its high performance and efficiency.
SCNet is a lightweight MSS model that compresses the representations of each stem while still achieving competitive results.
Similar to other band-split models~\cite{Luo2023, Lu2024}, SCNet operates in the frequency domain and separates a complex-valued input spectrogram into sub-bands.
However, it uses a fixed number of bands (low, middle, high) and the encoder compresses each band based on its sparsity of information to make the representations more uniform.
We train SCNet on Binaural-MUSDB with minimal modifications, using the default training parameters in the official GitHub repository\footnote{\tiny{\url{https://github.com/starrytong/SCNet/}}}~\cite{Tong2024} and the root mean squared error (RMSE) loss of the complex-valued spectrogram.

The only adjustment we made to the training paradigm of the binaural model is that we disable the scaling and channel flipping augmentations to maintain the spatial integrity of the training data.
The model is trained on 4 Nvidia Quadro RTX8000 GPUs for 130 epochs with an initial learning rate of 5e-4 using the Adam optimizer and a batch size of 4 audio segments per GPU.
We refer to this binaural model as Bi-SCNet and the stereo one as SCNet.

\vspace{-8pt}
\subsection{Perceptual Study}

Using the binaural and stereo models, we conducted an online perceptual study to evaluate how listeners judge the quality and spatial properties of the stems separated from the Binaural-MUSDB test mixtures.
Participants were instructed to complete the study using headphones in a quiet listening environment.
As a preliminary task, they reported the model of their headphones, tested their sound levels, verified their channel orientation, and completed a headphone screening test~\cite{Woods2017}.
Participants were asked to keep their volume fixed after setting it at a comfortable level, as all audio was normalized to -12 LUFS.
Only participants who correctly answered at least 5 of the 6 screening questions were included in the analysis.

The first part of the study measured perceived separation quality through pairwise comparisons~\cite{Cartwright2018, Ycart2020}.
For each trial, participants were told the target instrument group to evaluate: vocals, drums, or bass.
They were then presented with four audio clips: the input binaural mixture for context, the binaural reference stem, and two unlabeled model outputs: one from SCNet and one from Bi-SCNet.
Participants were asked to compare the outputs against the reference and select the version they believed was more similar to the reference, with a third option that both versions were equally similar.
The task consisted of several randomly selected pairwise comparisons plus a hidden reference in which the reference stem was presented as one of the candidate versions.
Participants who failed to select the reference in this trial were excluded from analysis.

The second part of the study evaluated perceived localization. 
Participants listened to several randomly selected stereo or binaural outputs and were asked to judge the spatial position of a specified instrument.
For each trial, they indicated their agreement on a 5-point Likert scale as to whether they heard the target instrument as coming from primarily a single direction.
If so, they selected the perceived direction by clicking within a semicircular spatial interface.
Next, participants similarly reported whether they heard the target instrument or other sounds coming from additional directions and, if applicable, marked up to two secondary locations on the interface.
Finally, they answered how spatially spread out the target sounded on a 5-point Likert scale with ``Very focused'' corresponding to a precise point-like location and ``Very spread out'' to a wide or smeared spatial image. 

The stimuli were chosen to cover both spatial and metric-driven conditions.
Each clip was 10-seconds long and was selected from the portion of the track with the peak RMS energy in the reference stem.
For simplicity, we studied separation in bass, vocals and drums, as the ``other'' group in the 4-stem MSS task has high variability in instrumentation across tracks.
For each instrument, we included examples where the target source was positioned on the left, right, or center of the binaural scene, allowing us to test whether metric behavior depended on source or source location.
Example stimuli are available online\footnote{\tiny{\url{https://richa-namballa.github.io/spacing-out-demo/}}}.

We recruited 30 participants for the perceptual study, the vast majority of whom were either audio researchers or engineers, or had musical training.
After filtering responses which failed screening or attention checks, our final data included responses from 26 participants: 96 pairwise comparisons and 132 localizations with at least 3 responses per condition.

\vspace{-8pt}
\subsection{Metrics}\label{subsec:metrics}

We examine the metrics from \cite{Namballa2025}, along with alternative GCC-PHAT variants proposed to improve the robustness of the TDOA estimation.
These include partial spectral whitening through PHAT-$\beta$~\cite{Donohue2007} and noise-based binary masking~\cite{Lee2020}.
We investigate these alternatives to assess how timing-based spatial measurements behave on separated binaural music stems.

\textbf{Interaural Cues.} $\Delta\text{ILD}$ is defined as the absolute difference in ILD values between the estimated ($\hat{\mathbf{s}}$) and reference ($\mathbf{s}$) stems, measured in decibels (dB).
\vspace{-4pt}
\begin{equation}\label{eq:delta_ild}
    \Delta\text{ILD} = |\text{ILD}(\mathbf{s}) - \text{ILD}(\hat{\mathbf{s}})|
\end{equation}

The ILD is calculated over the full signal duration as the ratio of the total energy in the left and right channels, where $N$ represents the total number of samples in the signal.
\vspace{-4pt}
\begin{equation}\label{eq:ild}
    \text{ILD}(\mathbf{s}) = 10 \cdot \log_{10} \left(\frac{ \sum_{n=0}^{N-1}{s_L[n]^2}}{ \sum_{n=0}^{N-1}{s_R[n]^2}}\right)
\end{equation}

Similiarly, $\Delta$ITD is defined as the difference in ITD values between the estimated and reference stems, measured in microseconds ($\mu$s).

\vspace{-4pt}
\begin{equation}\label{eq:delta_itd}
    \Delta\text{ITD} = |\text{ITD}(\mathbf{s}) - \text{ITD}(\hat{\mathbf{s}})|
\end{equation}

To estimate the ITD (more generally known as TDOA), the input signal $\mathbf{s}$ is segmented into 0.5-second frames using a Hann window function.
For each frame, the cross-correlation vector $\mathbf{c(s)}$ is computed between the left and right channels via GCC-PHAT~\cite{Knapp1976}.
Let $\mathcal{F}$ represent the Discrete Fourier Transform (DFT), $\mathcal{F}^{-1}$ its inverse, $*$ the complex conjugate, and $\odot$ the element-wise product.

\vspace{-4pt}
\begin{equation}\label{eq:gccphat}
    \mathbf{c(s)} = \mathcal{F}^{-1}\left(\frac{\mathcal{F}(\mathbf{s}_L) \odot \mathcal{F}(\mathbf{s}_R)^*}{|\mathcal{F}(\mathbf{s}_L) \odot \mathcal{F}(\mathbf{s}_R)^*|}\right)
\end{equation}

The ITD for the frame is then identified as the time lag corresponding to the maximum peak in $\mathbf{c(s)}$.
The final ITD for a two-channel signal is the mode of the lags across frames, weighted by the frames' energies.

\textbf{Weighted GCC-PHAT.} Traditional GCC-PHAT fully whitens the cross-power spectrum by normalizing by magnitude.
The PHAT-$\beta$ transform generalizes this step by adjusting the denominator in Equation~\ref{eq:gccphat} to ${\left|\mathcal{F}(\mathbf{s}_L)\odot\mathcal{F}(\mathbf{s}_R)^*\right|^\beta}$, where $\beta < 1$ partially preserves magnitude information before the TDOA estimation~\cite{Donohue2007}.

\begin{table*}[h!]
\centering
\vspace{-6pt}
\caption{Median separation and spatial error results on the test set of Binaural-MUSDB. The best results are in \textbf{bold}.}
\vspace{-6pt}
\resizebox{\textwidth}{!}{
\begin{tabular}{@{}lcccccccccccc@{}}
\toprule
\multirow{2}{*}{\bf Model}
& \multicolumn{3}{c}{\bf SRR (dB) $\uparrow$}
& \multicolumn{3}{c}{\bf SSR (dB) $\uparrow$}
& \multicolumn{3}{c}{\bf $\Delta$ITD ($\mu$s) $\downarrow$}
& \multicolumn{3}{c}{\bf $\Delta$ILD (dB) $\downarrow$} \\
\cmidrule(lr){2-4}
\cmidrule(lr){5-7}
\cmidrule(lr){8-10}
\cmidrule(lr){11-13}
& Bass & Drums & Vocals
  & Bass & Drums & Vocals
  & Bass & Drums & Vocals
  & Bass & Drums & Vocals \\
\midrule

Bi-SCNet
    & \bf{14.74} & \bf{15.00} & \bf{11.24}
    & {9.97} & \bf{10.74} & \bf{9.42}
    & \bf{476.19} & \bf{0.00} & {294.78}
    & \bf{0.06} & \bf{0.08} & \bf{0.11} \\

SCNet
    & {8.26} & {10.87} & {5.97}
    & \bf{10.00} & {10.03} & {8.38}
    & \bf{476.19} & \bf{0.00} & {385.49}
    & {0.38} & {0.37} & {0.48} \\
    
Demucs
    & {8.90} & {10.58} & {4.37}
    & {9.13} & {10.39} & {8.70}
    & \bf{476.19} & \bf{0.00} & \bf{0.00}
    & {0.20} & {0.31} & {0.42} \\

\bottomrule
\end{tabular}
}
\label{tab:all_metrics}
\vspace{-12pt}
\end{table*}

The partial whitening of the cross-power spectrum in the GCC-PHAT-$\beta$ algorithm was found to increase the robustness of the delay estimation in noisy and reverberant  environments~\cite{Donohue2007}.
However, this implementation has primarily been used for speech processing and, to the best of our knowledge, has not been extensively tested on musical signals.
We explore using this method in our alternative approach to ITD estimation in Section \ref{sec:itd}.

\textbf{Masked GCC-PHAT.}
The work in \cite{Lee2020} explores a method of overcoming diffuse noise and reverberation in the TDOA estimation between two microphones.
They use a binary mask to only include reliable frequency components in the GCC-PHAT calculation.
The discarded frequency components are chosen based on the coherent-to-diffuse power ratio (CDR) to identify the bins that are a result of diffuse noise and reverberation rather than direct sound.
Given that the Binaural-MUSDB data is synthesized from only head-related impulse responses (HRIRs) and does not include additional room acoustics other than the original effects from the Stereo-MUSDB stems, we opted for a simpler criteria to filter out noisy frequency components.
We binarized our mask as
\vspace{-6pt}
\begin{equation}
    M(m) = 
\begin{cases}
    0 & \text{if } |R(m)| < \delta_{T} \\
    1 & \text{otherwise}
\end{cases}
\end{equation}

where $|R(m)|$ is the energy (in dB) of the $m$-th frequency component of the cross-power spectrum $R$ and $\delta_{T}$ is the energy threshold.
The mask is applied to the normalized cross-power spectrum before inverting the signal back to the time domain.
\vspace{-6pt}
\begin{equation}
        \mathbf{c(s)} = \mathcal{F}^{-1}\left(\frac{\mathbf{M} \odot \mathcal{F}(\mathbf{s}_L) \odot \mathcal{F}(\mathbf{s}_R)^*}{|\mathcal{F}(\mathbf{s}_L) \odot \mathcal{F}(\mathbf{s}_R)^*|}\right)
\end{equation}

\textbf{Energy-based Ratios.} The approach described in \cite{Watcharasupat2024}, defines the total error between $\mathbf{s}$ and $\mathbf{\hat{s}}$ as
\vspace{-6pt}
\begin{equation}
    \mathbf{e}_\text{total} = \mathbf{s} - \mathbf{\hat{s}} = \mathbf{e}_\text{spat} + \mathbf{e}_\text{resid}
\end{equation}
These decompositions are estimated through a least-squares optimization of gain and delay to project $\mathbf{{s}}$ onto $\mathbf{\hat{s}}$, deriving the vector $\mathbf{\Tilde{s}}$.

SSR and SRR are computed as median frame-wise energy ratios in dB.
$\text{SSR}(\hat{\mathbf{s}};\mathbf{s}) = 10\log_{10}\!\left(\|\mathbf{s}\|^2 / \|\mathbf{e}_{\text{spat}}\|^2\right)$ measures the energy of the reference signal relative to the spatial error, while $\text{SRR}(\hat{\mathbf{s}};\mathbf{s}) = 10\log_{10}\!\left(\|\tilde{\mathbf{s}}\|^2 / \|\mathbf{e}_{\text{resid}}\|^2\right)$ measures the energy of the spatially corrected reference relative to the residual error.

\vspace{-8pt}
\section{Results}
\begin{figure}[ht]
    \centering
    \vspace{-6pt}
    \includegraphics[width=0.95\linewidth]{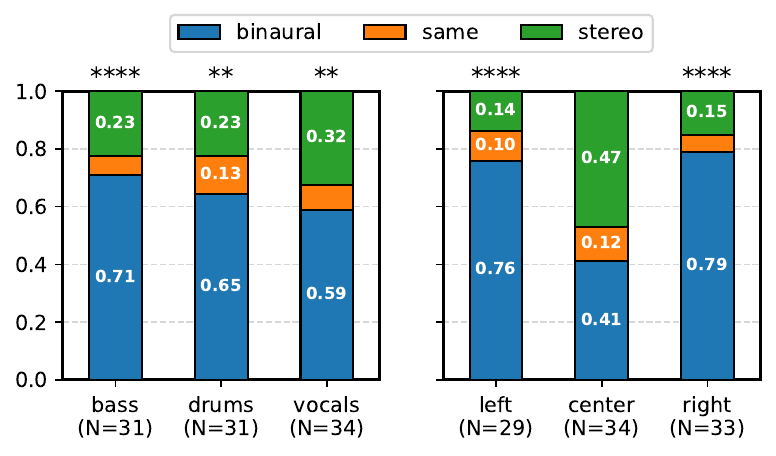}
    \vspace{-8pt}
    \caption{Vote proportion in pairwise comparisons between models by instrument (left) and source position (right). Asterisks indicate significant deviations from an equal preference distribution according to a $\chi^2$ test.}
    \label{fig:preferences}
    \vspace{-12pt}
\end{figure}

We begin with the results of the perceptual study and analyze the pairwise comparisons of the model outputs in their perceived similarity to the binaural reference.
Our goal is to identify whether participants could perceive a difference between the stereo and binaural models in terms of separation quality and spatial preservation of the sources.
As shown in Figure~\ref{fig:preferences}, participants more often selected the binaural model output over the stereo one across all instrument groups, with the difference in distribution being highly significant, according to a $\chi^2$ goodness-of-fit test.

Surprisingly, there is no significant difference in preference for the binaural model output in cases where the source is positioned in the center of the spatial image (directly in front of the listener).
Centered sources produce more similar binaural and stereo images, so we believe that the outputs of the two models may be harder to distinguish perceptually.
In contrast, sources positioned to the left or right require  stronger interaural cues to localize.
Therefore, the binaural model may preserve more spatial information which led participants to perceive the outputs as more similar to the reference.

\begin{figure}[ht]
    \centering
    \vspace{-8pt}
    \includegraphics[width=0.9\linewidth]{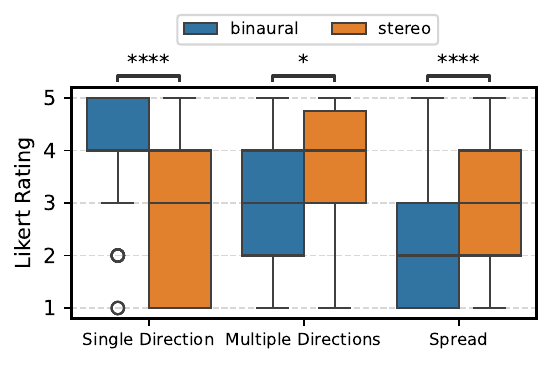}
    \vspace{-8pt}
    \caption{Likert scale ratings of perceived spatial artifacts and source spread. Asterisks indicate significant differences between models according to a Mann-Whitney U test~\cite{MannWhitney1947}. Agreement that target source was heard from a single direction (left). Agreement that target source or other sounds were heard from additional directions (center). Perceived spatial spread of the target source, where higher values indicate a wider or more smeared spatial image (right).}
    \label{fig:spatialness}
    \vspace{-12pt}
\end{figure}

The localization task results provide further support for the binaural model's utilization of spatial information. Outputs from the binaural model were perceived as more spatially focused, with lower described spread and fewer reports of sounds leaking from additional directions (Figure~\ref{fig:spatialness}).
These significant differences suggest that the binaural model better preserves spatial cues and introduces fewer spatial artifacts in the separation, offering a reliable foundation for evaluating the objective metrics.

We also examined inter-rater agreement by analyzing the comparisons for which there was a majority agreement in the chosen model output.
We found that in the 73.08\% of the conditions, a majority of participants selected the same option in the comparison, suggesting that listener preferences were generally consistent in the task.

\begin{figure}[ht]
    \centering
    \vspace{-8pt}
    \includegraphics[width=0.95\linewidth]{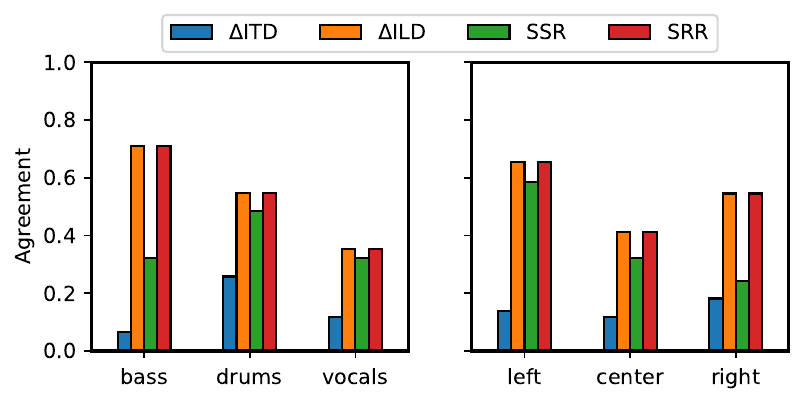}
    \vspace{-10pt}
    \caption{Fraction of pairwise comparisons for which each metric selected the same model as the listeners. Agreement is shown separately by instrument (left) and source position (right).}
    \label{fig:agreement}
    \vspace{-10pt}
\end{figure}

In Table \ref{tab:all_metrics}, we report the metrics for evaluating binaural MSS  and include results from Demucs~\cite{Rouard2023} for comparison with \cite{Namballa2025}.
Consistent with the advantages of in-domain training, Bi-SCNet outperforms the stereo models across the majority of sources and evaluation metrics.
To better understand the metrics' correlation with perception, we compare participants' responses against the rankings produced by each metric.
Following the analysis in \cite{Ycart2020}, we consider a metric and a participant response as agreeing if they both rank the models the same.

In Figure \ref{fig:agreement}, we see the overall agreement level is moderate, but that some metrics reflect listener preferences much better than others.
$\Delta$ILD and SRR align more consistently with ratings, while SSR occasionally nears their agreement levels but sometimes performs worse than as if the participants chose randomly (0.33).
$\Delta$ITD is an outlier with much less agreement across all instruments and positions.

The largest disparity in agreement between metrics occurs with  bass instruments.
Agreement is highest for bass sources among the level- and residual-error-based metrics ($\Delta$ILD and SRR), suggesting that the models had clearer differences related to residual leakage or level imbalance in the separated bass stems.
In contrast, agreement is lowest for the timing and spatial distortion metrics ($\Delta$ITD and SSR) for bass, indicating that either these metrics capture characteristics that are irrelevant to listeners' perceived quality or that the metrics are less stable when computed on separated low-frequency, narrow-bandwidth stems.

Furthermore, we observe that agreement is low for sources placed in the center of the binaural mixture.
This finding ties in to the earlier results from Figure~\ref{fig:preferences}: listeners have a harder time discriminating between the stereo and binaural outputs for centered sources.
Upon closer inspection of the vocal subset, we also find that the overall weak agreement for the vocals is driven primarily by the sources placed in the center.
In this condition, listeners more often selected the stereo-trained model, whereas most objective metrics favored the binaural-trained model.
This divergence suggests that the low agreement level is due to a specific failure mode of centered vocals, where the perceptual differences between binaural and stereo renderings are subtle, and that the metrics may reward properties that do not necessarily translate into better perceived quality.
We suspect that there may also be an effect of familiarity for listeners.
Since lead vocals are commonly presented in the center of commercial stereo mixes, participants are inherently more accustomed to centered stereo rendering and thus perceive it as more ``correct,'' even if the reference stem is binaural.

\vspace{-12pt}
\section{Interaural Time Difference Estimation}\label{sec:itd}

The perceptual study shows that $\Delta$ITD has the weakest agreement with listener preferences. Considering that $\Delta$ITD is intended to reflect a core binaural spatial cue, its poor alignment with perception motivates a closer review of how the metric behaves on separated music stems. 
Upon further investigation of the results, we find that for many estimated stems, especially the bass, the TDOA calculation often returns an ITD of 0$\mu$s regardless of the source's azimuth location (thus, explaining the identical $\Delta$ITD values across models in Table \ref{tab:all_metrics}).
This inaccuracy means that the metric can collapse to a centered estimate even when the source itself is not centered in the spatial image.

Furthermore, this failure is consistent with the known limitations of GCC-PHAT-based delay estimation methods~\cite{Donohue2007, Clifford2013, Lee2020}.
The GCC-PHAT algorithm is intended for broadband signals, as it relies on phase information across the frequency spectrum to find the peak correlation between channels.
The method can be sensitive to high frequency artifacts, as it discards magnitude in the calculation, making even minor, low-energy noise be weighted equally with the dense energy of the low frequency band.
A low-pass filter would not improve the calculation due to the introduction of inaudible phase distortion in the stopband.
With a narrow-band source such as the bass, the algorithm appears to work for the reference stem, but struggles to find the true TDOA of the estimated stems due to the cross-correlation peak centering at a lag of 0 samples.

In our ITD estimation calculations, we opted for a Hann window as suggested by \cite{Clifford2013} rather than the Tukey window used in \cite{Namballa2025}, but did not find a significant improvement in the stability of the calculation.
Likewise, trying a window longer than 0.5s did not impact the results.
In an attempt to correct the accuracy of the ITD estimation for binaural MSS, we explore two alternative strategies for TDOA estimation.
It should be noted that the failure occasionally occurs in the estimated vocals.
However, for the remainder of this section, we focus on the most severe case: bass.

\vspace{-6pt}
\subsection{Weighted GCC-PHAT}

Theoretically, the GCC-PHAT-$\beta$ variation should address the noise issue in our estimated bass stems by incorporating magnitude information in the calculation. In reality, it brings to light a very interesting trade-off: robustness vs. accuracy.

\begin{figure}[h]
    \vspace{-8pt}
    \centering
    \begin{subfigure}{0.49\columnwidth}
        \centering
        \includegraphics[width=\linewidth]{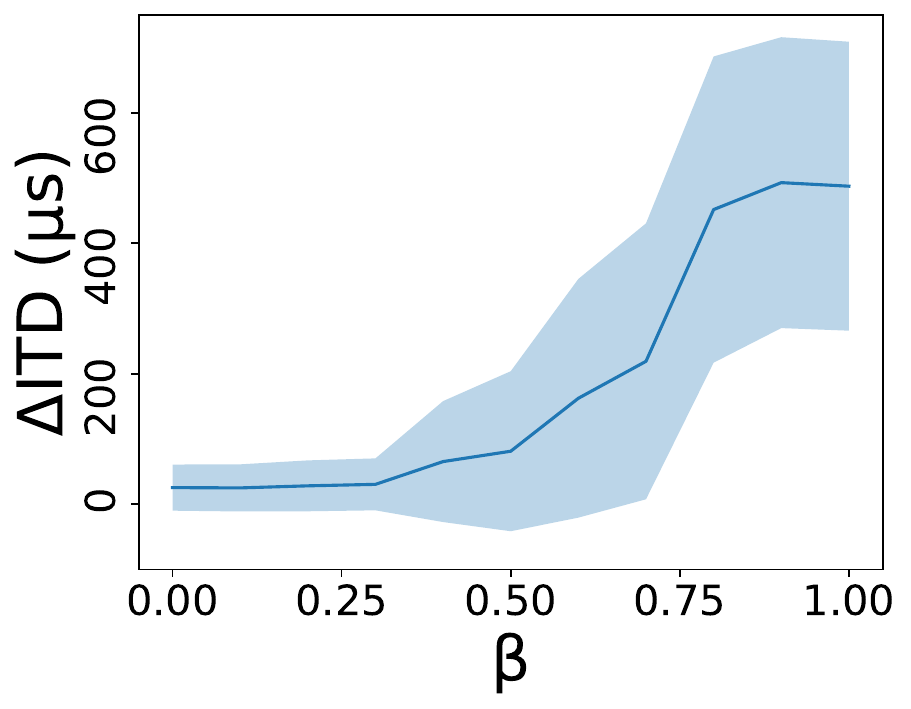}
        \caption{Bass}
    \end{subfigure}
    \hfill
    \begin{subfigure}{0.44\columnwidth}
        \centering
        \includegraphics[width=\linewidth]{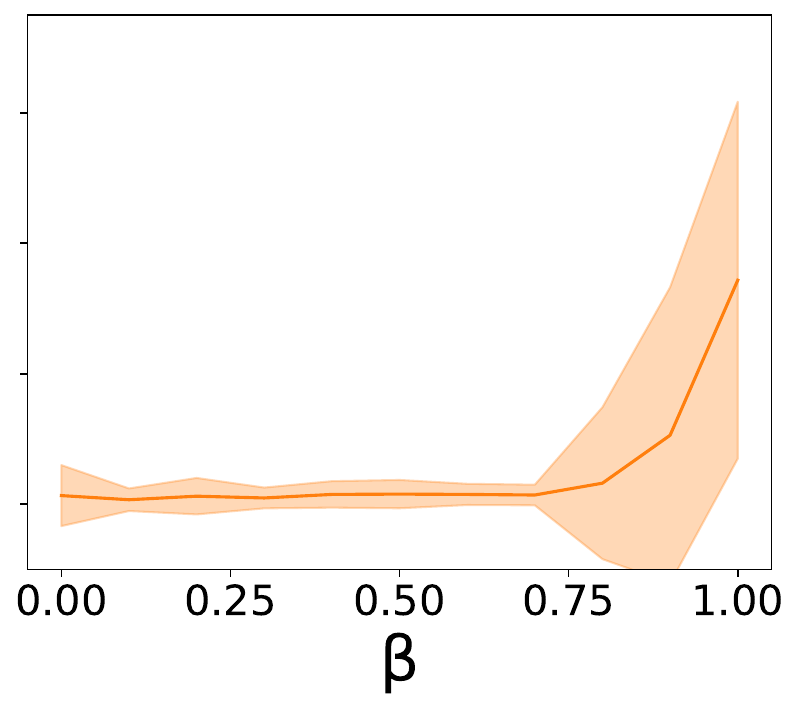}
        \caption{Vocals}
    \end{subfigure}
    \vspace{-6pt}
    \caption{Mean $\Delta$ITD error of the estimated stems from Bi-SCNet for the test set of Binaural-MUSDB computed with GCC-PHAT-$\beta$.}
    \label{fig:weight_simulation}
    \vspace{-8pt}
\end{figure}

A key decision in the use of GCC-PHAT-$\beta$ is the tuning of the $\beta$ parameter.
Using $\beta$ = 1 is equivalent to the full phase transform, while $\beta$ = 0 corresponds to generalized cross correlation.
In Figure~\ref{fig:weight_simulation}, we see how the values of $\Delta$ITD for the test set's bass stems change as $\beta$ varies.
We find that the error is minimized between the reference and estimated stems when $\beta$ = 0 (traditional GCC) and most varied with $\beta$ = 1 (fully whitened PHAT).
In contrast, the same calculations on the vocal stems are much less sensitive to changes in the weighting factor.
These results indicates that using $\beta$ = 0 would provide the most robust TDOA estimation for computing our metrics.
However, our focus on the estimated stems operates under the assumption that the computed ITDs of the reference bass stems are correct.
To verify that the reference ITD values are indeed the ground truths when using GCC-PHAT-$\beta$, we compare the computed ITD values with the ITD measurements modeled by the HRIRs which were used to synthesize Binaural-MUSDB~\cite{Armstrong2018}.
We found that varying $\beta$ when calculating the HRIRs' ITDs does not significantly alter their values.
Nonetheless, as we reduce the amount of whitening in the phase transform (decrease $\beta$), the computed ITD values of the reference stems begin to drift away from those of the HRIRs.
Using $\beta = 1$ provides the most accurate ITD estimation, directly conflicting with the previously established desire for robustness.
This trend highlights the fragile trade-off: increasing the stability of the ITD calculation by decreasing $\beta$ will also lower the accuracy of the computation.

\subsection{Noise Invariance}

\begin{figure}[h]
    \vspace{-8pt}
    \centering
        \centering
        \includegraphics[width=0.9\linewidth]{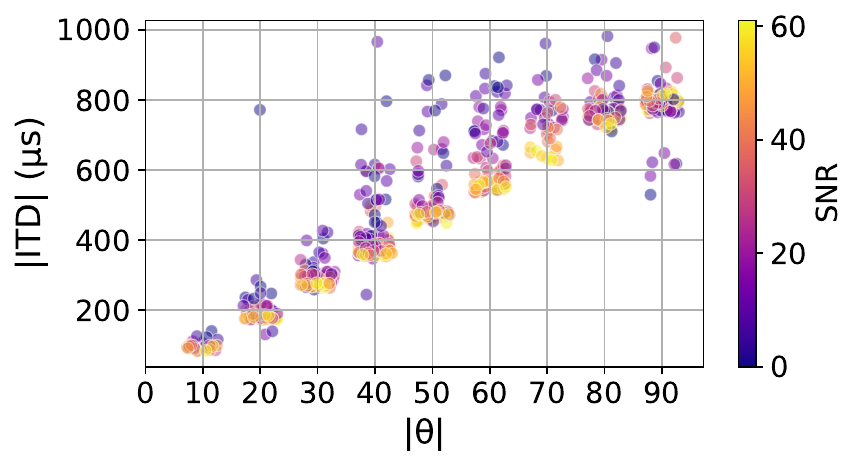}
    \vspace{-8pt}
    \caption{ Distribution of ITD values for the bass reference stems of Binaural-MUSDB across angles, varied by SNR.}
    \label{fig:noise_by_azimuth}
    \vspace{-8pt}
\end{figure}

To better understand the GCC-PHAT calculation's sensitivity to noise, we looked into how the ITD values vary across different SNRs for vocals, drums, and bass.
We simulated various noise levels by adding uncorrelated noise to each channel of the reference stems and found that, above a very low SNR level, the ITD estimations are noise invariant for vocals and drums.
However, there was a clear failure in the bass ITD calculations around an SNR of 45dB, which reflects an almost imperceptible noise level.
This finding suggests that even near-perfect source separation would still result in unreliable ITD estimation, indicating that GCC-PHAT is unsuitable for bass.
Looking even closer at the bass ITD values, we see a clear trend between the variance of the ITD calculations and the angle of the azimuth location in Figure \ref{fig:noise_by_azimuth}.
For lower SNR values, the farther the source is located from the origin (higher $|\theta|$), the more the ITD estimations deviate, demonstrating the instability of ITD calculations for sources not located directly in front of the listener.

\begin{figure}
    \centering
    \vspace{-6pt}
    \includegraphics[width=0.9\linewidth]{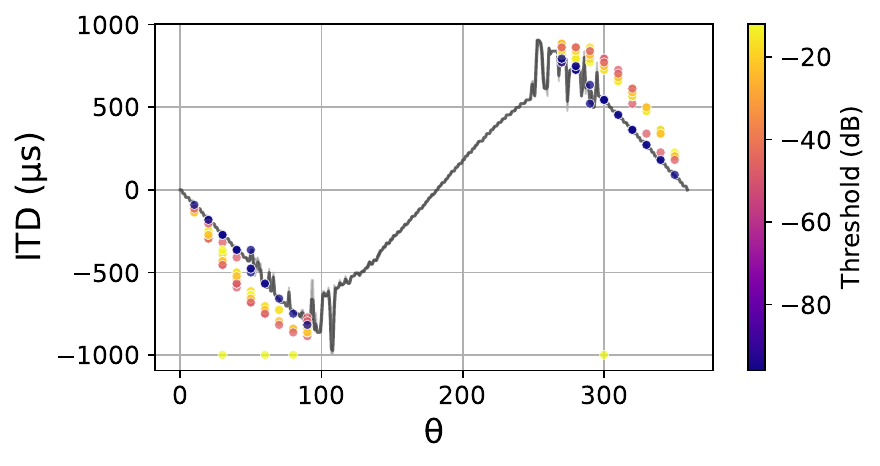}
    \vspace{-6pt}
    \caption{ITD estimates of the Binaural-MUSDB bass reference stems with the ITD curve of the HRIRs used to synthesize them for varying $\delta_T$ thresholds.}
    \vspace{-12pt}
    \label{fig:mask_gccphat_hrir}
\end{figure}

\vspace{-12pt}
\subsection{Masked GCC-PHAT}

Since the primary issue with using GCC-PHAT on the estimated bass stems is the presence of noise and artifacts, we explore using Masked GCC-PHAT to remove the noisy frequency components from the ITD calculation.
This approach should ``zero-out'' the phase of low energy components in the cross-power spectrum with the assumption that the majority of those bins originate from noise, interference, and artifacts of the separation.
These signals contain phase information that impedes the correct phase estimation needed for an accurate ITD calculation.
Upon applying this approach to the test set with varying thresholds, we see from the drift away from the ITD curve in Figure \ref{fig:mask_gccphat_hrir} that we run into a similar issue as before: the robustness of the calculation increases, but the accuracy decreases.

\vspace{-12pt}
\section{Conclusion}
We investigated how well the metrics used to evaluate binaural MSS models correlate with human judgment.
Our results demonstrate that the agreement between the metrics and perception varies depending on position and instrument, and that the metrics are not equally perceptually relevant.
Level- and residual-error-based metrics ($\Delta$ILD, SRR) are more aligned with human preference than timing and spatial-distortion ones ($\Delta$ITD, SSR).
Furthermore, we tested the fragility of the GCC-PHAT calculation in estimating $\Delta$ITD, a core indicator of spatial quality, and identified that there is a robustness-accuracy tradeoff, especially in the case of bass instruments.
We plan to expand this work by conducting an expanded perceptual study with larger-capacity MSS models more tailored towards binaural music, exploring the behavior of instruments beyond the 4-stem scenario, and identifying more metrics intended to evaluate spatial integrity in the task of binaural MSS.

\section{Ethics Statement}

Participants of the perceptual study were unpaid volunteers who provided informed consent at the time of participation.
They could revoke their consent at any point during or after the experiment.
The study was conducted with oversight by the New York University (NYU) Institutional Review Board (IRB) through IRB-FY2020-4157 and with the consultation of Prof. Pablo Ripollés of the Department of Psychology, Faculty of Arts \& Science, at NYU.

\bibliography{ISMIR2026}

\end{document}